\documentclass{naturemod}
\usepackage{graphicx}
\usepackage{epsfig}
\usepackage[font={small,sf}]{caption}
\def \ling {merging}

\def \quantum {state}
\def \sab {S(AB)}
\def \sa {S(A)}
\def \sb {S(B)}
\def \sacb {S(A|B)}
\def \sbca {S(B|A)}
\def \Ic {I(A\rangle B)}

\def \R {R}

\def \tc {\bar{T}}
\newcommand{\beq}{\begin{equation}}
\newcommand{\eeq}{\end{equation}}
\newcommand{\beqa}{\begin{eqnarray}}
\newcommand{\eeqa}{\end{eqnarray}}
\newcommand{\beqar}{\begin{eqnarray*}}
\newcommand{\eeqar}{\end{eqnarray*}}

\def \s {\,\,\,\,}

\def \ra {\rangle}

\def \r {\rho}
\def\rab{\rho_{AB}}
\def \ra {\rho_A}
\def \rb {\rho_B}

\def \X {X}
\def \Y {Y}

\def \ratea {R_A}
\def \rateb {R_B}
\def \ratet {R_T}
\def \eass {E_A}
\newcommand{\proj}[1]{\ket{#1}\bra{#1}}
\newcommand{\bra}[1]{\langle #1 |}
\newcommand{\ket}[1]{| #1 \rangle}
\renewcommand{\captionlabelfont}{\bf}
\begin{document}

\title{Quantum information can be negative}

\author{Michal Horodecki$^1$, Jonathan Oppenheim $^2$ \& Andreas Winter$^3$}
\twocolumn[
\begin{@twocolumnfalse}
\maketitle

\begin{affiliations}
 \item Institute of Theoretical Physics and Astrophysics, University of Gda\'nsk, 80--952 Gda\'nsk, Poland
 \item Department of Applied Mathematics and Theoretical Physics, University of Cambridge, Cambridge CB3 0WA, U.K.
 \item Department of Mathematics, University of Bristol, Bristol BS8 1TW, U.K.
\end{affiliations}
\date{{\it Submitted Feb. 20th, 2005}}

\vspace{12pt}

\begin{abstract}
  Given an unknown quantum state distributed over two systems, we determine how much 
  quantum communication is needed to transfer the full state to one system.  
  This communication measures the \emph{partial information} one system needs
  conditioned on it's prior information. It turns out to be given
  by an extremely simple formula, the \emph{conditional entropy}. 
  In the classical case, partial information must always
  be positive, but we find that in the quantum world this physical
  quantity can be negative.
  If the partial information is positive, its sender
  needs to communicate this number of quantum bits to the receiver;
  if it is negative, the sender and receiver instead \emph{gain} the corresponding potential for
  future quantum communication.
  We introduce a primitive \emph{quantum state merging}
  which optimally transfers partial information.
  We show how it enables a systematic understanding of 
  quantum network theory, and discuss several important applications
  including distributed compression,  multiple
  access channels and multipartite assisted entanglement distillation (localizable entanglement).  
Negative channel capacities also receive 
a natural interpretation.
\end{abstract}
\end{@twocolumnfalse}
]

\vspace{3in}
'Ignorance is strength' is one of the three cynical mottos of Big Brother
in George Orwell's {\it 1984}. Most of us would naturally incline to the
opposite view, trying continually to increase our knowledge on just about
everything. But regardless of preferences, we are thus confronted by two
questions: how much is there to know? And, how large
is our ignorance in a given situation?

The reader will observe that the formulation of these questions addresses
the \emph{quantity} of information, not its content, and
this is simply because the latter is hard to assess and to compare. The
former approach to classical information was pioneered by
Claude Shannon\cite{Shannon1948},
who provided the tools and concepts to scientifically answer the first of our two
questions: the amount of information originating from a source is the
memory required to faithfully represent its output. For the case of a
statistical source, on which we will concentrate throughout, this amount
is given by its \emph{entropy}.

To approach the second question, let us introduce a two-player game. One
participant (Bob) has some incomplete prior information $Y$,
the other (Alice) holds some missing information $X$:
we think of $X$ and $Y$ as random variables, and Bob has
prior information due to possible correlations between $X$ and $Y$.
If Bob wants to learn $X$,
how much additional information does Alice need to send him?  
This is one of the key problems of classical information theory,
since it describes a ubiquitous scenario in information networks.
It was solved by Slepian and 
Wolf\cite{slepian-wolf} who proved
that the amount of information that Bob needs is given by a
quantity called the \emph{conditional entropy}.
It measures the partial information that Alice must send to Bob so that he gains 
full knowledge of $\X$ given his previous knowledge from $\Y$, and it is just the difference 
between the entropy of $(X,Y)$
taken together (the total information) and the entropy of $\Y$ (the prior
information).
Of course, this partial information is always a positive quantity.
Classically, there would be no meaning to negative information.

In the quantum world, the first of our two 
questions, how to quantify quantum information,
was answered by Schumacher\cite{Schumacher1995}, who 
showed that the minimum number of quantum bits required
to compress quantum information 
is given by the quantum (von Neumann) entropy.
To answer the second question, let us now consider the quantum version of the
two-party scenario above:
Alice and Bob each possess a system in some 
unknown quantum state with the total density operator being $\rab$
and each party having states with density operators $\ra$ and $\rb$ respectively.
The interesting case is where Bob is correlated with
Alice, so that he has some prior information about her state. 
We now ask how much additional 
quantum information Alice needs to send him, so that 
he has the full state (with density operator $\rab$).  
Since we want to quantify the quantum partial information, 
we are interested in the minimum amount of
quantum communication to do this, allowing unlimited 
classical communication -- the latter type
of information being far easier to transmit than the former as it can 
be sent over a telephone while the
former is extremely delicate and must be sent using a special quantum channel.

Since we are interested in informational quantitities, we go to the limit of many copies of
state $\rab$ and vanishing but non-zero errors in the protocol.
We find here that the amount of partial quantum information that Alice needs to send Bob 
is given by the quantum conditional entropy, which is exactly 
the same quantity as in the classical
case but with the Shannon entropy changed to the
von Neumann entropy: 
\beq
\sacb\equiv\sab-\sb \s ,
\eeq
where $\sb$ is the entropy of Bob's state $\rb$ and $\sab$ is the
entropy of the joint state $\rab$.
For quantum states, the conditional entropy
can be negative\cite{Wehrl78,HH94-redun,cerfadami}, and 
thus it is rather surprising that this quantity has a physical interpretation
in terms of how much quantum communication is needed to gain 
complete quantum information i.e., possession of a system in the total state $\rab$.

However, in the above scenario, the negative
conditional entropy can be clearly interpreted.  We find 
that when $\sacb$ is negative,
Bob can obtain the full state using only classical communication,
and additionally, 
Alice and Bob will have the potential to transfer additional quantum
information in the future at no additional cost to them.  Namely, they end
up sharing $-\sacb$ Einstein-Podolsky-Rosen (EPR) pairs\cite{EPR},
i.e. pure maximally entangled states
$\frac{1}{\sqrt{2}}(\ket{00}+\ket{11})$, which can be used
to teleport\cite{teleportation} quantum 
states between the two parties using only classical communication.
Negative partial information thus also
gives Bob the potential to receive future quantum information for free. 
The conditional entropy plays the same role in quantum information 
theory as it does
in the classical theory, except that here, the quantum conditional 
entropy can be negative in
an operationally meaningful way. 
One could say that the ignorance of Bob, the conditional
entropy, if negative, precisely cancels the amount by which
he knows too much\cite{Schroedinger}; the latter being
just the potential future communication gained.

This solves the well known puzzle of how to interpret
the quantum conditional entropy which has persisted despite interesting attempts
to understand it\cite{cerfadami}.  Since there are no conditional probabilities
for quantum states, $\sacb$ is not an entropy as in the classical case. But by
going back to the definition of information in terms of storage space needed
to hold a message or state, one can make operational sense of this quantity.

Let us now turn to the protocol which allows Alice to transfer 
her state to Bob's site in the above scenario
(we henceforth adopt the common usage 
of refering directly to manipulations on ``states'' -- meaning a
manipulation on a physical system in some quantum state).
We call this {\it quantum state merging}, since Alice is effectively merging
her state with that of Bob's. 
Let us recall that   
in quantum information theory, faithful 
state transmission means that 
while the state merging protocol may depend on 
the density operator of the source, it must
succeed with high probability for any pure state sent.
An equivalent and elegant way of expressing this criterion is to imagine 
that $\rab$ is part of a pure state 
$\ket\psi_{ABR}$, which includes a reference system $\R$.  
Alice's goal is to transfer the
state $\ra$ to Bob, and we demand that after the protocol,
the total state still has high fidelity with $\ket\psi_{ABR}$
(meaning they are nearly identical);
see Figure 1 which includes a high-level description of the protocol.
\begin{figure}
  \centering
  \includegraphics[width=8cm]{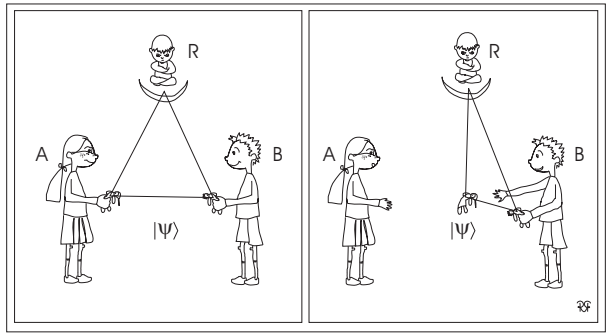}
  \caption{Diagrammatic representation of the process of state merging.
           Initially the state $\ket{\psi}$ is shared between the three systems
           R(eference), A(lice) and B(ob). After the communication Alice's system is in a
           pure state, while Bob holds not only his but also her initial
           share. Note that the reference's state $\rho_R$ has
           not changed, as indicated by the curve separating
           R from AB.
The protocol for state merging is as follows:
Let Alice and Bob have a large number $n$ of the state $\rab$.
To begin, we note that we only need to describe the protocol for
negative $\sacb$, as otherwise Alice and Bob can share $n\sacb$ EPR
pairs (by sending this number of quantum bits) and create
a state $\ket{\psi}_{AA'BB'R}$ with $S(AA'|BB')<0$.  This is because
adding an EPR pair reduces the conditional entropy
by one unit.
However, $\sacb<0$ is equivalently expressed as
$S(A) > S(AB) = S(R)$, and it is known\cite{shor-cap,Lloyd-cap,igor-cap} that 
measurement in a uniformly random basis on Alice's $n$ systems projects Bob
and R into a state $\ket{\varphi}_{BR}$ whose reduction to R
is very close to $\rho_R$. But this means that Bob can, by a local
operation, transform $\ket{\varphi}_{BR}$ to $\ket{\psi}_{ABR}$.
Finally, by coarse-graining the random measurement, Alice
essentially projects onto a good quantum code\cite{shor-cap,Lloyd-cap,igor-cap}
of rate $-\sacb$; this still results in Bob obtaining the full
state $\rho_{AB}$, but now, just under $-n\sacb$ EPR pairs are also created.
These codes can also be obtained by an alternative
construction\cite{DevetakWinter-hash}.
}
\end{figure}
The essentially element of
state merging is that $\rho_R$ must be unchanged, and Alice must decouple
her state from $R$.  This also means (seemingly paradoxically) that as far as any outside party is
concerned, neither the classical nor quantum communication is coupled
with the merged state.

\noindent
Let us now consider three instructive and simple examples:  
\begin{enumerate}
  \item Alice has a completely unknown state which we can
    represent as the maximally mixed density matrix 
    $\rho_A=\frac{1}{2}(\proj{0}_A+\proj{1}_A)$, and Bob has
    no state (or a known state $\ket{0}_B$). In this case, 
    $\sacb=1$ and Alice must send one qubit down the quantum
    channel to transfer
    her state to Bob.  She could also send half of an EPR state 
    to Bob, and use quantum teleportation\cite{teleportation}
    to transfer her state.

  \item The classically correlated state $\rab=\frac{1}{2}(\proj{00}_{AB}+\proj{11}_{AB})$.
    We imagine this state as being part of a pure state with the reference
    system $R$,
    $\ket{\psi}_{ABR}=\frac{1}{\sqrt{2}}(\ket{0}_A\ket{0}_B\ket{0}_R
                                         +\ket{1}_A\ket{1}_B\ket{1}_R)$.
    In this case, $\sacb = 0$, and
    thus no quantum information needs to be sent. Indeed,
    Alice can measure her state in the
    basis $\ket{0}\pm\ket{1}$, and inform Bob of the result.
    Depending on the outcome of the measurement,
    Bob and $R$ will share one of two states
    $\ket{\phi^{\pm}}_{BR}=\frac{1}{\sqrt{2}}(\ket{0}_B\ket{0}_R\pm\ket{1}_B\ket{1}_R)$,
    and by a local operation, Bob can always transform the state to 
    $\ket{\psi}_{A'BR}=\frac{1}{\sqrt{2}}(\ket{0}_{A'}\ket{0}_B\ket{0}_R
                                          +\ket{1}_{A'}\ket{1}_B\ket{1}_R)$ with $A'$ being an ancilla at Bob's site.
    Alice has thus managed to send her state to Bob, while fully preserving
    their entanglement with $R$.

  \item For the state
    $\ket{\phi^{+}}_{AB}=\frac{1}{\sqrt{2}}(\ket{0}_A\ket{0}_B+\ket{1}_A\ket{1}_B)$,
    $\sacb=-1$, and Alice and Bob can keep this
    shared EPR pair to allow future transmission of quantum 
    information, while Bob creates the EPR pair 
    $\ket{\phi^{+}}_{A'B}$ locally.  I.e. transferring a pure 
    state is trivial since the pure state is known and can be created locally.
\end{enumerate}

Let us now make a couple of observations about state merging.
First, the amount of classical communication that is required
is given by the number of quantum codes in Alice's projection:
the quantum mutual information
$I(A:R)=S(A)+S(R)-S(AR)$ between Alice and the reference $\R$.  
Secondly, the measurement of Alice makes her state completely
product with $\R$, thus reinforcing the interpretation 
of quantum mutual information 
as the minimum entropy production of any local
decorrelating process\cite{GroismanPW04,huge}.
This same quantity is also equal to the amount of
irreversibility of a cyclic process: Bob initially has a state, 
then gives Alice her share
(communicating $S(A)$),
which is finally merged back to him (communicating $\sacb$). 
The total quantum communication of this cycle is $I(A:R)$ quantum bits.

Because state merging is such a basic primitive, it allows 
us to solve a number of other
problems in quantum information theory fairly easily.
We now sketch four particularly striking applications.

\noindent {\bf Distributed quantum compression}:
for a single party, a source emitting states with density matrix 
$\ra$ can be compressed at a rate
given by the entropy $\sa$ of the source by performing quantum 
data compression\cite{Schumacher1995}.
Let us now consider the distributed scenario -- we 
imagine that the source emits states with density matrix 
$\r_{A_1A_2...A_m}$, and distributes it over $m$ parties.
The parties wish to compress their shares as much as possible 
so that the full state can be reconstructed by a 
single decoder. Until now the general solution of this problem has
appeared intractable\cite{adhw2004}, but it becomes very simple once we allow
classical side information for free, and use state merging.

Remarkably, the parties can compress the state at the 
total rate $S(A_1A_2\ldots A_m)$ -- the Schumacher limit\cite{Schumacher1995}
for collective
compression -- even though they must operate seperately.
This is analogous to the classical result, the 
Slepian-Wolf theorem\cite{slepian-wolf}.
We describe the quantum solution for two parties and depict
the rate region in Figure 2.  
\begin{figure}
  \label{fig:sw}
  \centering
  \includegraphics[width=8cm]{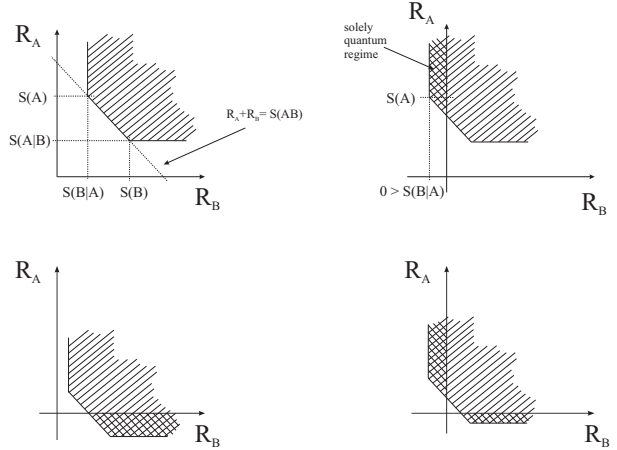}
  \caption{The rate region for distributed compression by two parties
    with individual rates $R_A$ and $R_B$. The total rate $R_{AB}$ is bounded by $\sab$.
    The top left diagram shows the rate region of a source with
    positive conditional entropies; the top right and bottom left
    diagrams show the purely quantum case of sources
    where $\sbca<0$ or $\sacb<0$. It is even possible that
    both $\sbca$ and $\sacb$ are negative, as shown in the
    bottom right diagram, but observe that the rate-sum $S(AB)$
    has to be positive.
If one party compresses at a rate $\sb$, then the other party can
over-compress at a rate $\sacb$, by merging her state with the state which will end up with
the decoder.  Time-sharing gives the full rate region, since the
bounds evidently cannot be improved.
Analogously, for $m$ parties $A_i$, and all
subsets $T\subseteq\{1,2,\ldots,m\}$ holding a combined
state with entropy $S(T)$, the rate sums $\ratet = \sum_{i\in T} R_{A_i}$
have to obey
$  \ratet > S(T|\tc) $
with $\tc = \{1,2,\ldots,m\}\setminus T$ the complement of set $T$.
}
\end{figure}

\noindent {\bf Noiseless coding with side information}: 
related to distributed compression is the case where only Alice's state needs
to arrive at the  decoder, while Bob can send part of his state to the decoder in order to help Alice lower
her rate. The classical case of this problem was introduced
by Wyner\cite{wyner-side}. For the quantum case, we demand that the full state $\rab$ be preserved in the protocol, but do not place
any restriction on what part of Bob's state may be at the decoder and what part can remain with him.
For one-way protocols, we find using state merging that if $\ra$ and $\rb$ are encoded at rates 
$R_a$ and $R_b$ respectively, then the decoder
can recover $\ra$ if and only if $R_a\geq S(A|U)$ and $R_b\geq E_p(AU:R)-S(A|U)$
with $R$ being the purifying reference system, $U$ being a system with its state produced by some quantum channel 
on $\rb$, and $E_p(AU:R)\equiv\min_\Lambda S(A\Lambda(U))$ being the entanglement of purification\cite{IBMHor2002}.
The mimimum is taken over all channels $\Lambda$ acting on $U$.

\noindent {\bf Quantum multiple access channel}:
in addition to the central questions of information theory we asked earlier,
{\it how much is there to know, and how great is our ignorance}, information
theory also concerns itself with communication rates.  In the quantum world,
the rate at which quantum information can be sent down a noisy channel is related to
the coherent information $\Ic$ which was previously defined as\cite{SchumacherNielsen}
$\max\{\sb-\sab,0\}$.
This quantity is the quantum counterpart of Shannon's mutual information;
when maximized over input states,
it gives the rate at which quantum information can be sent from Alice to
Bob via a noisy quantum channel\cite{shor-cap, Lloyd-cap, igor-cap}.  
As with the classical conditional entropy,
Shannon's classical mutual
information\cite{Shannon1948} is always positive, and
indeed it makes no sense to have classical channels with negative capacity.
However, the relationship between the coherent information and the quantum 
channel capacity contained a puzzle. As the latter was thought to
be meaningful as a positive quantity, the former
was defined as the maximum of $0$ and $\sb-\sab$ since it
could be negative.

We will see that negative values \emph{do} make sense,
and thus propose that $\Ic$ should not be defined as 
above, but rather as $\Ic = -\sacb$.
It turns out that negative capacity, impossible in classical information theory
has its interpretation in a situation with two senders.

We imagine that Alice and Bob wish to send independent quantum 
states to a single decoder Charlie
via a noisy channel which acts on both inputs. This 
problem is considered by Yard \emph{et al.}\cite{ydh2005}.
Our approach using \quantum\ \ling\ provides a
solution also when either of the
channel capacities are negative, and gives the following 
better achievable rates:
\beqa
\ratea&\leq& I(A\rangle CB)\s, \nonumber\\
\rateb &\leq& I(B\rangle CA)\s, \nonumber\\ 
\ratea+\rateb &\leq& I(AB\rangle C)\s,
 \label{eq:capregion}
\eeqa
where $\ratea$ and $\rateb$ are the rates of Alice and Bob for sending quantum states.
Here, we use our redefinition of the coherent information, in that we allow it to be
negative.  In achieving these rates, one party can send (or invest) $I(A\rangle C)$
quantum bits
to merge her state with the decoder.
The second party then already has Alice's state at the decoder,
and can send at the higher rate $I(B\rangle AC)$.
This provides an interpretation of negative channel capacities: if the channel of one party
has negative coherent information,
this means that she has to invest this amount of entanglement
to help her partner achieve the highest rate.  The protocol is for one of the parties
to merge their state with the state held by the decoder.  
The expressions in (\ref{eq:capregion}) are in formal analogy
with the classical multiple access channel.

\noindent {\bf Entanglement of assistance (localizable entanglement)}:
consider Alice, Bob and $m-2$ other parties sharing
(many copies of) a pure quantum state.  
The entanglement of assistance $\eass$\cite{entass}
is defined as the maximum entanglement that the other 
parties can create between Alice and Bob by local measurements
and classical communication.  For many parties, this is often referred
to as {\it localizable entanglement}\cite{vpc2004}, although here we work
in the regime of many copies of the shared state.
This problem was recently solved for up to four parties\cite{svw2005}, and
can be generalized to an arbitrary number $m$ of parties
using state merging (using universal
codes depending only on the density matrix of the helper,
as described in Figure 1). 
We find that 
the maximal amount of entanglement that
can be distilled between Alice and Bob, with the help
of the other parties, is
given by the minimum entanglement across any bipartite cut of the system
which separates Alice from Bob:
\beq  
  \eass = \min_{T} \{ S(AT) , S(B\overline{T}) \}
\eeq
where the minimum is taken over all possible partitions of the 
other parties into groups $T$ and its complement $\overline{T}=\{1,\ldots,m-2\}\setminus T$.
To achieve this, each party in turn merges their
state with the remaining parties, preserving the minimum cut entanglement.

\medskip
We have described a fundamental quantum information
primitive, \emph{state merging}, and demonstrated some of its many
applications. There are also conceptual implications.
For example, the celebrated strong subadditivity
of quantum entropy\cite{lieb:ruskai}, $S(A|BC) \leq S(A|B)$,
receives a clear interpretation and transparent proof: having more prior information
makes state merging cheaper.
Our results also shed new light on the foundations
of quantum mechanics: it has long been known that there are
no conditional probabilities, so defining conditional
entropy is problematic. Just replacing classical entropy with quantum
entropy gives a quantity which can be positive or negative.
Quite paradoxically, only the negative part was understood
operationally, as quantum channel capacity, which, if anything,
made the problem even more obscure.
State merging "annihilates" these problems with each other.
It turns out that the 
puzzling form of quantum capacity as a conditional entropy
is just the flip-side of our
interpretation of quantum conditional entropy as partial
quantum information, which makes equal sense
in the positive and negative regime. The key point is to realize that
in the negative regime, one can gain entanglement \emph{and} transfer
Alice's partial state, while in the positive regime, only the partial state
is transfered.

On a last note, 
we wish to point out that remarkably and despite the
formal analogy, the classical scenario does not occur as a classical
limit of the quantum scenario -- we consider both classical and
quantum communication, and there is no meaning to preserving
entanglement in the classical case.
We have only just begun to grasp the full implications of state
merging and negative partial information;
a longer technical account\cite{how-merge2}, with rigorous proofs
and further applications is in preparation.
 
\bibliography{localize}


\begin{addendum}
  \item[Acknowledgements]
    This work was done at the Isaac Newton Institute (Cambridge)
    August-December 2004 and we are grateful for the institute's hospitality.
We thank Bill Unruh for his helpful comments on an earlier draft of this paper,
and Roberta Rodriquez for drawing Figure 1.  
We acknowledge the support of EC grants RESQ, No. IST-2001-37559, QUPRODIS, No. IST-2001-38877, 
PROSECCO (IST-2001-39227),
MH was additionally supported by the Polish Ministry of Scientific
Research and Information Technology under grant No.
PBZ-MIN-008/P03/2003, JO was additionally supported by
the Cambridge-MIT Institute and the Newton Trust, and 
AW was additionally supported by the U.K. Engineering and Physical 
Sciences Research Council. 

  \item[Competing Interests] The authors declare that they have no
    competing financial interests.
  \item[Correspondence] Correspondence and requests for materials
    should be addressed to Jonathan Oppenheim.
    (Email: {\tt j.oppenheim (at) damtp.cam.ac.uk})

\end{addendum}

\end{document}